\documentclass[a4paper]{article}

\usepackage{siunitx}
\usepackage{multirow}
\usepackage{tabularx}
\usepackage{booktabs}
\usepackage{textgreek}
\usepackage{upgreek}
\usepackage{INTERSPEECH2022}

\title{Refining DNN-based Mask Estimation using CGMM-based EM Algorithm \\ for Multi-channel Noise Reduction}
\name{Julitta Bartolewska, Stanisław Kacprzak, Konrad Kowalczyk\thanks{This research received financial support from the National Science Centre of Poland under grant number DEC-2017/25/B/ST7/01792 and from the Foundation for Polish Science under grant number First TEAM/2017-3/23 (POIR.04.04.00-00-3FC4/17-00) which is co-financed by the European Union under the European Regional Development Fund.}}

\address{AGH University of Science and Technology, Institute of Electronics, 30-059 Krakow, Poland}
\email{\{bartolew, skacprza, konrad.kowalczyk\}@agh.edu.pl}

\begin{document}

\maketitle
\begin{abstract}
In this paper, we present a method that allows to further improve speech enhancement obtained with recently introduced Deep Neural Network (DNN) models. We propose a multi-channel refinement method of time-frequency masks obtained with single-channel DNNs, which consists of an iterative Complex Gaussian Mixture Model (CGMM) based algorithm, followed by optimum spatial filtration. We validate our approach on time-frequency masks estimated with three recent deep learning models, namely DCUnet, DCCRN, and FullSubNet. We show that our method with the proposed mask refinement procedure allows to improve the accuracy of estimated masks, in terms of the Area Under the ROC Curve (AUC) measure, and as a consequence the overall speech quality of the enhanced speech signal, as measured by PESQ improvement, and that the improvement is consistent across all three DNN models.
\end{abstract}

\section{Introduction}
\label{sec:intro}

Quality and intelligibility of the speech signal can be greatly affected in the presence of undesired background noises. It results in performance degradation of numerous voice-based technologies. Hence, the task of noise reduction is of great importance and aims at extracting the source signal from the noisy mixture.

Traditional spatial filtering techniques include Minimum Variance Distortionless Response (MVDR) beamforming \cite{Veen1988} and multi-channel Wiener filtering \cite{Brandstein2001}, which require estimation of the second-order statistics (SOS). Popular statistical modelling estimators are the Minimum Mean Square Error (MMSE) \cite{Hendriks2010}, Maximum Likelihood (ML) \cite{Higuchi2016, Higuchi2017}, and Maximum A Posteriori (MAP) \cite{Higuchi2017}, based on which time-frequency masks can be estimated under the assumption of CGMM using iterative Expectation-Maximization (EM) algorithm \cite{Higuchi2016, Higuchi2017}.

In recent years, Deep Neural Networks (DNNs) have been widely used for single- and multi-channel speech enhancement. Due to their ability to model nonlinear relationships, they have attracted broad attention. They can be broadly divided into time and time-frequency domain methods, both for single- and multi-channel cases. The time domain methods, referred to as waveform mapping, try to map the noisy speech to clean speech directly \cite{Rethage2018wavenet, Pandey2019tcnn}. Recent improvement in multi-channel methods that operate in time domain is reported in \cite{Pandey2021tparn}. On the other hand, for the time-frequency methods usually a learning target is defined as clean speech spectrogram or desired masking (i.e  Ideal Ratio Mask). Many methods focus only on magnitude features, due to the problems with estimation of phase information (phase of noisy signal is re-used in enhanced signal), however methods that try to reconstruct phase, i.e  with complex Ideal Ratio Mask (cIRM) \cite{Williamson2015complex}, increase in popularity \cite{dcunet,dcrn,fullsubnet}. For the case of multi-channel speech enhancement additional spatial filtering can be performed with the use of MVDR \cite{Erdogan2016improved, Heymann2016neural}, spatial information can be directly used as a features \cite{Deng2020dnn} or indirectly in end-to-end methods \cite{Tolooshams2020channel}. 

In our work, we are interested in combining single-channel deep neural network processing with additional spatial post-processing and further improving overall multi-channel noise reduction, an approach investigated also in \cite{Erdogan2016improved,Zhang2017}. In this paper, we propose to refine time-frequency masks obtained from single-channel DNNs with an iterative CGMM-based ML algorithm \cite{Higuchi2016,Higuchi2017} followed by spatial filtering for improved multi-channel noise reduction.
This kind of post-processing helps to leverage spatial features and obtained results show that it is able to consistently improve performance of speech enhancement. 
With respect to the EM algorithm in~\cite{Higuchi2016}, we introduce CGMM weights, and contrary to \cite{Higuchi2017} we assume their time-dependency, while in contrast to \cite{Matsui2018}, clustering of spatial features is not involved. Our work extends the approach in \cite{Erdogan2016improved} by introducing the mask refinement procedure, which however differs from \cite{Zhang2017} where the enhanced speech is iteratively fed back into the single-channel DNN mask estimator in an attempt to refine the spectral masks. Finally, with respect to~\cite{MartinDonas2020}, our approach does not involve Kalman filtering to model time dependencies.

\section{Proposed method}
\label{sec:method}

\subsection{Problem formulation and system overview}
\label{ssec:problem}

We consider a scenario with a single speaker recorded using an $M$-element microphone array, $m\in{\big\{1, \ldots, M\big\}}$, in a noisy environment. The vector of microphone observations in the Short-Time Fourier Transform (STFT) domain is given by
\begin{equation}
\mathbf{y}^{(t,f)}=\mathbf{s}^{(t,f)}+\mathbf{n}^{(t,f)} \; ,
\end{equation}
where subscripts $t$ and $f$ denote time frame and frequency indices, respectively, $\mathbf{s}^{(t,f)}$ and $\mathbf{n}^{(t,f)}$ denote the speech and noise signals as observed at the microphones, where $\mathbf{y}^{(t,f)}=[y^{(t,f,1)}, \ldots, y^{(t,f,M)}]^{\textrm T}$, $\mathbf{s}^{(t,f)}=[s^{(t,f,1)}, \ldots, s^{(t,f,M)}]^{\textrm T}$, and $\mathbf{n}^{(t,f)}=[n^{(t,f,1)}, \ldots, n^{(t,f,M)}]^{\textrm T}$. In this work, we aim to extract the target speech signal $s^{(t,f,1)}$ by performing multi-channel filtering of noisy microphone observations $\mathbf{y}^{(t,f)}$ using the  time-frequency (T-F) dependent multi-channel filter $\mathbf{w}_{\mathrm{MC}}^{(t,f)}$ (defined similarly to $\mathbf{y}^{(t,f)}$), which can be written as 
\begin{equation}
\hat{s}^{(t,f,1)}={\mathbf{w}_{\mathrm{MC}}^{(t,f)}}^{\mathrm{H}}\mathbf{y}^{(t,f)} \; .
\end{equation}

\begin{figure}[!t]
\centerline{\includegraphics[width=0.9\columnwidth]{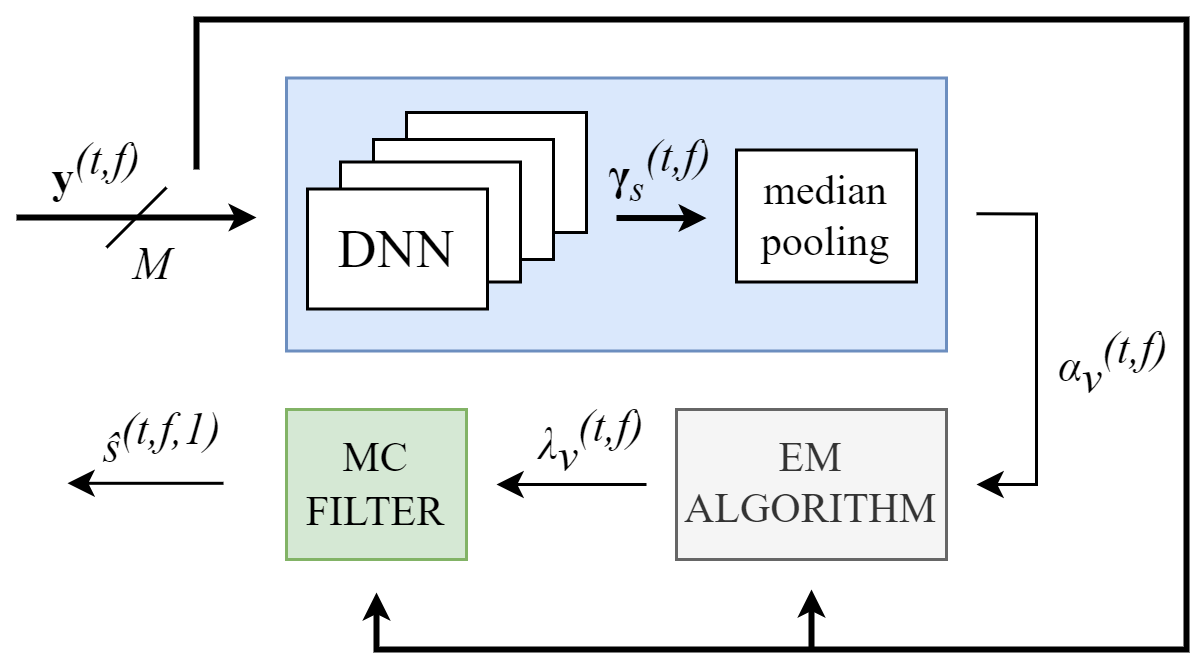}}
\caption{The proposed framework with mask refinement.}
\label{fig:scheme}
\vspace{-4.4mm}
\end{figure}

The proposed method for multi-channel noise reduction consists of the following processing steps, which are schematically depicted in Figure \ref{fig:scheme}. We begin by computing time-frequency (T-F) masks using a deep neural network. In essence, any DNN model that infers either real or complex T-F masks can be used for this purpose, and for robustness such inference should be performed at each microphone. Then, we convert these masks into energetically constrained real T-F masks, so that they can represent speech prior. They are next subjected to median pooling, before they undergo further refinement. In the next step, the obtained intermediate T-F masks are processed using the proposed mask refinement procedure which exploits spatial and spectral information inherent in the microphone signals, as well prior knowledge on speech presence from the previous step. The refinement is performed by the presented CGMM-based EM algorithm. The posterior probabilities of the EM algorithm represent final source and noise mask estimates, based on which the second-order statistics required for the presented multi-channel filter are computed.

\subsection{DNN-based estimation of intermediate T-F masks}
\label{ssec:DNN}
As an initial processing stage, we propose to estimate T-F masks using a deep neural network which operates on a single microphone channel. In principle, any network which produces real or complex T-F mask for the source or noise could be used.

In this work, we consider three DNN models which produce the Complex Ratio Masks (CRMs), namely the DCUnet \cite{dcunet} (the network architecture is widely used, however the original ICLR paper is retracted by authors due to an error in data processing step), DCCRN \cite{dcrn} and FullSubNet \cite{fullsubnet}. DCUnet uses U-Net \cite{Ronneberger2015u} based complex-valued building blocks to process spectrograms, and estimate a complex ratio mask (with polar coordinate-wise complex-valued masking method). The network is trained to optimize weighted Source-to-Distortion Ratio (wSDR) loss. DCCRN, also operates on T-F representation and combines the advantages of DCUnet, Convolution Recurrent Network (CRN) \cite{Tan2018convolutional} and Long Short-Term Memory (LSTM) \cite{Weninger2015speech} (with convolutional and recurrent layers operating in the complex domain). Similarly to \cite{Luo2019conv}, the network optimizes Scale-Invariant Signal-to-Noise Ratio (SI-SNR) loss. The third network, FullSubNet, incorporates a fusion of full-band and sub-band models to produce single-channel speech enhancement. Full-band model is able to capture the global context and cross-band dependencies, while sub-band model captures local spectral patterns. The network is trained to optimize complex Ideal Ratio Mask (cIRM).

Next, we convert the DNN-based masks to the real-valued T-F masks, constrained to their ability to preserve speech energy, as they will be needed in the multi-channel refinement step (Sec. \ref{ssec:EM}). Irrespective of whether single-channel (SC) filtration is performed using the real or complex masks (i.e., real gain is used in SC Wiener filtering, while complex CRM is inferred from the considered DNNs  \cite{dcunet,dcrn,fullsubnet}), the energetically constrained real T-F mask $\gamma_{s}^{(t,f,m)}$ for the $m$-th microphone can be computed as
\begin{equation}
\label{eq:gamma_m}
\gamma_{s}^{(t,f,m)} =  \frac{|\xi^{(t,f,m)}|^{2}}{|y^{(t,f,m)} - \xi^{(t,f,m)}|^{2} + |\xi^{(t,f,m)}|^{2}} \;,
\end{equation}
where $\xi^{(t,f,m)}$ is an estimate of the complex source spectrum given by $\xi^{(t,f,m)} =  \big(\Re{\{H^{(t,f,m)}\}} \Re{\{y^{(t,f,m)}\}} - \Im{\{H^{(t,f,m)}\}} \Im{\{y^{(t,f,m)}\}} \big) + 
\big(\Re{\{H^{(t,f,m)}\}} \Im{\{y^{(t,f,m)}\}} + \Im{\{H^{(t,f,m)}\}} \Re{\{y^{(t,f,m)}\}} \big)$, where $H^{(t,f,m)}$ is the filter coefficient obtained from DNN, and $\Re$ and $\Im$ denotes real and imaginary parts, respectively.

For increasing robustness, we propose to perform such a DNN-based mask estimation and subsequent filtration \eqref{eq:gamma_m} for each microphone of the array, which yields $\mathbf{\upgamma}_{s}^{(t,f)} = \bigl[\gamma_{s}^{(t,f,1)}, \ldots, \gamma_{s}^{(t,f,M)}\bigr]$. Then, we perform median pooling as
\begin{equation}
\alpha_{s}^{(t,f)} = \mathrm{Med}(\mathbf{\upgamma}_{s}^{(t,f)}) \; ,
\end{equation}
and additionally compute T-F masks for noise as
\begin{equation}
\alpha_{n}^{(t,f)} = 1 - \alpha_{s}^{(t,f)} \; .
\end{equation}

\subsection{Refining T-F masks with CGMM-based EM algorithm}
\label{ssec:EM}
In order to refine the T-F masks $\alpha_{v}^{(t,f)}$ obtained after median pooling, we propose to exploit the spatial properties inherently included in the recorded microphone signals. The proposed mask refinement procedure employs a two-component Complex Gaussian Mixture Model (CGMM) \cite{Higuchi2016,Higuchi2017} that models the multi-channel microphone signal in the T-F domain using zero-mean complex multivariate Gaussian distribution, which can be written as
\begin{equation}
\label{eq:y}
\mathbf{y}^{(t,f)} \sim \sum_{v} \beta_{v}^{(t,f)} \mathcal{N}_{c}\left(0, \phi_{v}^{(t,f)} \mathbf{R}_{v}^{(f)}\right) \;,
\end{equation}
where $v \in \{s,n\}$ denotes speech presence or absence in the noisy mixture, respectively, $\beta_{v}^{(t,f)}$ denotes the weight of the $v$ mixture component and satisfies $\sum_{v}\beta_{v}^{(t,f)}=1$, $\phi_{v}^{(t,f)}$ denotes signal variance, and $\mathbf{R}_{v}^{(t,f)}$ is the spatial correlation matrix.

Different to the approach in~\cite{Higuchi2016,Higuchi2017}, we assume that the mixture weights are time-dependent. They are represented by the a priori speech presence (or absence) probability, and as approximation of these probabilities, we propose to use the T-F masks after median pooling, i.e. we assume that $\beta_{v}^{(t,f)} \approx \alpha_{v}^{(t,f)}$. Next, the parameters of the underlying CGMM model can be found with the Maximum Likelihood (ML) approach. Following derivations in~\cite{Higuchi2016,Higuchi2017}, the update rules for the E-step and the M-step of the resulting EM algorithm are given by
\begin{equation}
\lambda_{v}^{(t,f)} = \frac{\alpha_{v}^{(t,f)} \mathcal{N}_{c}\left(\mathbf{y}^{(t,f)} | 0, \phi_{v}^{(t,f)} \mathbf{R}_{v}^{(f)}\right)}{\sum_{v}\alpha_{v}^{(t,f)}\mathcal{N}_{c}\left(\mathbf{y}^{(t,f)} | 0, \phi_{v}^{(t,f)} \mathbf{R}_{v}^{(f)}\right)} \; ,
\end{equation}
\begin{equation}
\phi_{v}^{(t,f)} = \frac{1}{M} \operatorname{tr}\left(\mathbf{y}^{(t,f)} {\mathbf{y}^{(t,f)}}^{\mathrm{H}} {\mathbf{R}_{v}^{(f)}}^{-1}\right) \; ,
\end{equation}
\begin{equation}\label{eq:corr_update}
\mathbf{R}_{v}^{(f)} = \frac{1}{\sum_{t} \lambda_{v}^{(t,f)}} \sum_{t} \lambda_{v}^{(t,f)} \frac{1}{\phi_{v}^{(t,f)}} \mathbf{y}^{(t,f)} {\mathbf{y}^{(t,f)}}^{\mathrm{H}} \; ,
\end{equation}
where the posterior probability $\lambda_{v}^{(t,f)}$ corresponds to the refined T-F mask for $v \in \{s,n\}$. Note that in contrast to~\cite{Higuchi2017}, the proposed refining EM algorithm does not involve update rules for the time-dependent mixture weights, and thus they are kept fixed over iterations of the EM algorithm. 

Finally, note that initialization of $\mathbf{R}_{v}^{(f)}$ can be performed using \eqref{eq:corr_update} with $\lambda_{v}^{(t,f)} = \alpha_{v}^{(t,f)}$ and signal variance excluded from the equation, hence the normalization of the matrix is performed for each frequency by dividing the matrix by its trace averaged over the microphones. Additionally, for increasing the algorithm robustness in a single speaker case, first-rank approximation of the source correlation matrix $\mathbf{R}_{s}^{(f)}$ can be used.

\subsection{Multi-channel noise reduction filtering}
\label{ssec:filter}

Given the refined T-F masks for the source and noise, i.e., $\lambda_{s}^{(t,f)}$ and $\lambda_{n}^{(t,f)}$, we propose to perform noise reduction with the multi-channel Wiener filter given by the following closed-form solution
\begin{equation}
\mathbf{w}_{\mathrm{MC}}^{(t,f)} = \frac{{\mathbf{\Phi}_{\mathbf{n}}^{(f)}}^{-1} \; \mathbf{r}^{(f)}}{{\mathbf{r}^{(f)}}^{\mathrm{H}} \; {\mathbf{\Phi}_{\mathbf{n}}^{(f)}}^{-1} \; \mathbf{r}^{(f)}}\;\; \sqrt{\frac{\lambda_{s}^{(t,f)}}{\lambda_{s}^{(t,f)} + \lambda_{n}^{(t,f)}}} \; ,
\end{equation}
where noise and source second-order statistics are computed as
\begin{equation}
\mathbf{\Phi}_{\,\mathbf{n}}^{(f)} = \frac{1}{\sum_{t} \lambda_{n}^{(t,f)}} \sum_{t} \lambda_{n}^{(t,f)} \mathbf{y}^{(t,f)} {\mathbf{y}^{(t,f)}}^{\mathrm{H}} \; ,
\end{equation}
\begin{equation}
\mathbf{\Phi}_{\,\mathbf{y}}^{(f)} = \frac{1}{\sum_{t} \lambda_{s}^{(t,f)}} \sum_{t} \lambda_{s}^{(t,f)} \mathbf{y}^{(t,f)} {\mathbf{y}^{(t,f)}}^{\mathrm{H}} \; ,
\end{equation}
\begin{equation}
\mathbf{\Phi}_{\,\mathbf{s}}^{(f)} = \mathbf{\Phi}_{\,\mathbf{y}}^{(f)} - \mathbf{\Phi}_{\,\mathbf{n}}^{(f)} \; ,
\end{equation}
and $\mathbf{r}^{(f)}$ denotes the normalized steering vector obtained by performing singular value decomposition (SVD) of $\mathbf{\Phi}_{\,\mathbf{s}}^{(f)}$, taking eigenvector corresponding to the largest value and normalizing it with respect to the reference microphone.

\section{Experiments and result evaluation}
\label{sec:eval}

\subsection{Experimental setups, DNN models, and evaluation measures}
For experimental evaluation of the proposed framework, we consider a scenario in which a single speaker is recorded, in a mid-sized room in presence of background noise, using a nested 6-element microphone array (which comprises two 4-element sub-arrays with inter-microphone spacing of \SI{0.05} and \SI{0.15}{\meter}, respectively). The Room Impulse Responses (RIRs) are simulated using the image-source method \cite{Allen1979image,Habets2006rir} for rooms of random size drawn from the respective ranges {7-8}~\SI{}{\meter} $\times$ {5-6}~\SI{}{\meter} $\times$ {3-4}~\SI{}{\meter}, with a constant source-array distance of \SI{1}{\meter}, and randomly set Reverberation Time (RT) (ranging from \SI{0.2} to \SI{0.5}{\second}, achieved by adjusting wall absorption coefficients). Speech and noise signals are randomly drawn from the MUSAN database \cite{musan}, which includes speech recordings from LibriVox and noise signals from Free Sound and Sound Bible datasets. The microphone signals contain the reverberant speech and noise at four Signal-to-Noise Ratios (SNRs) of \SI{-5}, \SI{0}, \SI{5} and \SI{10}{\dB}. The sampling frequency amounts to \SI{16}{\kHz}, and we perform 512-point STFT with Hann window of \SI{32}{\ms} length and 50\% overlap. For frequencies below 1kHz, we process the signals of a smaller sub-array, while the larger sub-array is used above \SI{1}{\kHz}. Note that, as suggested in previous studies \cite{Higuchi2016,Higuchi2017}, we perform 20 iterations of the EM algorithm.

For the estimation of T-F masks using DNN, we use three networks described in Sec. \ref{ssec:DNN}, namely the  DCUnet~\cite{dcunet}, DCCRN~\cite{dcrn}, and FullSubNet~\cite{fullsubnet}. Regarding their implementations, we decided to use publicly available pre-trained models to enable straightforward reproduction of our experiments. Two of the models are based on the asteroid framework \cite{Pariente2020asteroid}, %
and their pre-trained models are available in the Hugging Face repository:
DCUnet\footnote{{https://huggingface.co/JorisCos/DCUNet\_Libri1Mix\_enhsingle\_16k}} and DCCRN\footnote{{https://huggingface.co/JorisCos/DCCRNet\_Libri1Mix\_enhsingle\_16k}} (all models were trained on LibriMix dataset \cite{Cosentino2020librimix}). For the FullSubNet network, we use the pre-trained model \footnote{{https://github.com/haoxiangsnr/FullSubNet}} provided by the authors (the model was trained with 2020 DNS Challenge dataset \cite{Reddy2020interspeech}). The network is capable to work in real-time using cumulative normalization \cite{Tan2018convolutional}, however, in our experiments we use it in batch-mode.

To evaluate the accuracy of the estimated time-frequency masks, we use the Receiver Operating Characteristic (ROC) curves which represent the True Positive Rate against the False Positive Rate, while as a single-value accuracy metric, we additionally compute their Areas Under Curves (AUCs). For the evaluation of noise reduction, the following measures are used: the Perceptual Evaluation of Speech Quality improvement (\textDelta PESQ) \cite{PESQ} between the enhanced and the reference input microphone, the Signal-to-Distortion Ratio (SDR) \cite{SDR}, Short-Time Objective Intelligibility (STOI) \cite{STOI}, and Normalized-Covariance Measure (NCM) \cite{NCM}. 

All results, presented for a given SNR level, are calculated by averaging over the results obtained for $1000$ test recordings with random source and noise signals, and random room setups.

\subsection{Results and discussion}

\begin{figure}[!t]
\centerline{\includegraphics[width=\columnwidth]{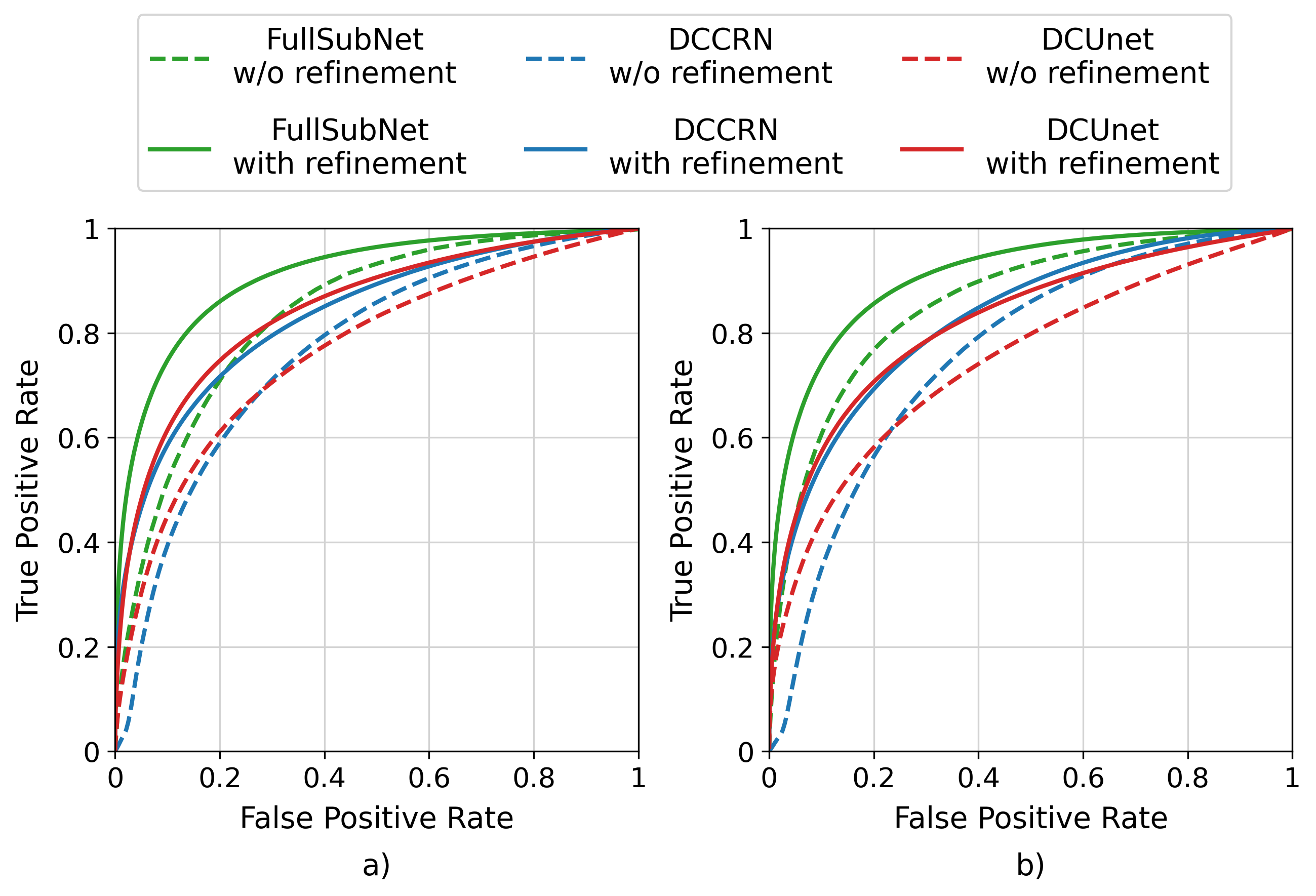}}
\vspace{-8pt}
\caption{Averaged Receiver Operating Characteristic (ROC) curves for noisy mixtures at input SNR of a) \SI{0}{\dB} and b) \SI{10}{\dB}.}
\label{fig:rocs}
\vspace{-2mm}
\end{figure}

\begin{figure}[!t]
\centerline{\includegraphics[width=0.95\columnwidth]{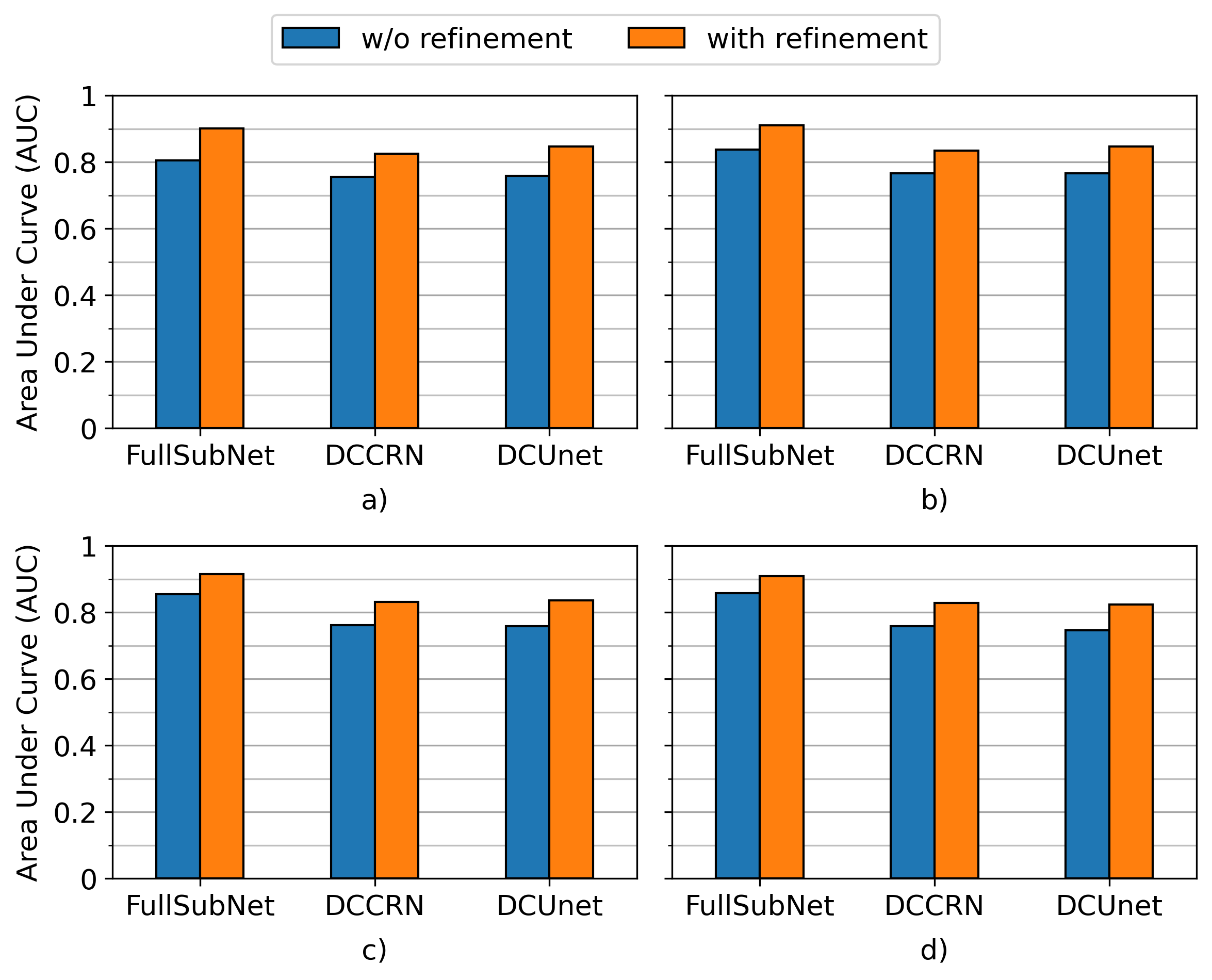}}
\vspace{-8pt}
\caption{Averaged Areas Under ROC Curves (AUCs) for noisy mixtures with input SNR equal to: a) \SI{-5}{\dB}, b) \SI{0}{\dB}, c), \SI{5}{\dB}, and d) \SI{10}{\dB}. The higher value the better result.}
\label{fig:aucs}
\vspace{-2mm}
\end{figure}

\begin{table*}[ht!]
\centering
\caption{Averaged performance measures for noisy mixtures with input SNR = \SI{0}{\dB}. The higher values the better results.}
\resizebox{0.8\textwidth}{!}{\begin{tabular}{cccc@{\hspace{-.5\arrayrulewidth}}ccc@{\hspace{-.5\arrayrulewidth}}cccc@{\hspace{-.5\arrayrulewidth}}cccc}%
 & & & & \multicolumn{3}{c}{\textbf{FullSubNet}} & & \multicolumn{3}{c}{\textbf{DCCRN}} & & \multicolumn{3}{c}{\textbf{DCUnet}} \\ \cmidrule(lr){5-7}
 \cmidrule(lr){9-11} 
 \cmidrule(lr){13-15} 
 & & refinement & & \multicolumn{1}{l}{CRM} & \multicolumn{1}{c}{SC} & \multicolumn{1}{c}{MC} & & \multicolumn{1}{c}{CRM} & \multicolumn{1}{c}{SC} & \multicolumn{1}{c}{MC} & & \multicolumn{1}{c}{CRM} & \multicolumn{1}{c}{SC} & \multicolumn{1}{c}{MC} \\ %
 \cmidrule(lr){1-3}
 \cmidrule(lr){5-7}
 \cmidrule(lr){9-11} 
 \cmidrule(lr){13-15} 
\multirow{2}{*}{\textbf{SDR {[}dB{]}}} & \multirow{8}{*}{} & w/o & & 9.3 & 8.3 & 9.2 & & 8.6 & 7.8 & 9.0 & & 8.3 & 7.5 & 8.9 \\ 
 & & with & & - & 9.3 & \textbf{9.4} & & - & 8.5 & \textbf{9.2} & & - & 8.9 & \textbf{9.1} \\ %
 \cmidrule(lr){1-3}
 \cmidrule(lr){5-7}
 \cmidrule(lr){9-11} 
 \cmidrule(lr){13-15} 
\multirow{2}{*}{\textbf{\textDelta PESQ}} & & w/o & & 0.61 & 0.36 & 0.66 & & 0.57 & 0.45 & 0.58 & & 0.47 & 0.22 & 0.60 \\
 & & with & & - & 0.58 & \textbf{0.84} & & - & 0.48 & \textbf{0.76} & & - & 0.52 & \textbf{0.82} \\ %
 \cmidrule(lr){1-3}
 \cmidrule(lr){5-7}
 \cmidrule(lr){9-11} 
 \cmidrule(lr){13-15} 
\multirow{2}{*}{\textbf{STOI}} & & w/o & & 0.83 & 0.81 & 0.82 & & 0.82 & 0.81 & 0.81 & & 0.81 & 0.79 & 0.82 \\ 
 & & with & & - & 0.85 & \textbf{0.86} & & - & 0.81 & \textbf{0.83} & & - & 0.85 & \textbf{0.86} \\ %
 \cmidrule(lr){1-3}
 \cmidrule(lr){5-7}
 \cmidrule(lr){9-11} 
 \cmidrule(lr){13-15} 
\multirow{2}{*}{\textbf{NCM}} & & w/o & & 0.76 & 0.75 & 0.76 & & 0.73 & 0.73 & 0.72 & & 0.68 & 0.67 & 0.70 \\ 
 & & with & & - & 0.81 & \textbf{0.83} & & - & 0.79 & \textbf{0.81} & & - & 0.77 & \textbf{0.80} \\ %
\end{tabular}}
\label{tab:final_res}
\end{table*}

Let us begin with an investigation of the proposed refinement procedure of the noise masks estimated using three studied DNN models. Figure \ref{fig:rocs} presents the ROC curves for input SNR of \SI{0} and \SI{10}{\dB}, which are calculated by averaging over the ROC curves \cite{phdthesisT,scikit-learn} obtained for 1000 recordings. The ROC curves for masks which underwent the proposed refinement procedure are always notably improved in comparison with the ROC curves for masks obtained directly from the respective DNN models. This improvement is particularly visible for low input SNR, where the refinement brings the highest increase in mask accuracy, effectively increasing true positives and reducing false positives. The AUC metric for all considered DNN models, with and without EM-based refinement, for various input SNR values is depicted in Figure \ref{fig:aucs}. These results show that the AUC is always increased by the proposed refinement, irrespective of the noise level and DNN model used for mask estimation. A close to constant improvement in AUC values observed for all input SNRs (shown in Figure \ref{fig:aucs}), combined with the increase in true positives and decrease in false positives (shown in Figure \ref{fig:rocs}), indicate that the proposed refinement procedure increases the accuracy of time-frequency masks.

Next, we investigate an effect of using the refined masks for multi-channel noise reduction. Table \ref{tab:final_res} presents the SDR, PESQ improvement, STOI, and NCM results obtained by the proposed multi-channel Wiener filtration (hereafter denoted as MC) using the refined masks and analogous processing without the refinement. For comparison, it also shows single-channel Wiener filtering (denoted as SC) using both refined and non-refined masks, as well as single-channel complex filtering using complex ratio masks (denoted as CRM) estimated by DNN models. As can be observed, mask refinement leads to an improvement in terms of all evaluation measures in case of single-channel and multi-channel Wiener filters, irrespective of the used DNN model (when comparing the impact of mask refinement alone for a given filter type). In particular, comparing all filtering approaches, the proposed multi-channel filtering based on the refined masks achieves the best results for all evaluation metrics and any selected DNN model. Based on the reference results of CRM, we conclude that single-channel filtering using complex DNN masks outperforms the single-channel Wiener filter that is based on a real mask bound to 1 (such that it can be used as prior probability). However, it fails to outperform the proposed approach. 

\begin{figure}[!h]
\centerline{\includegraphics[width=\columnwidth]{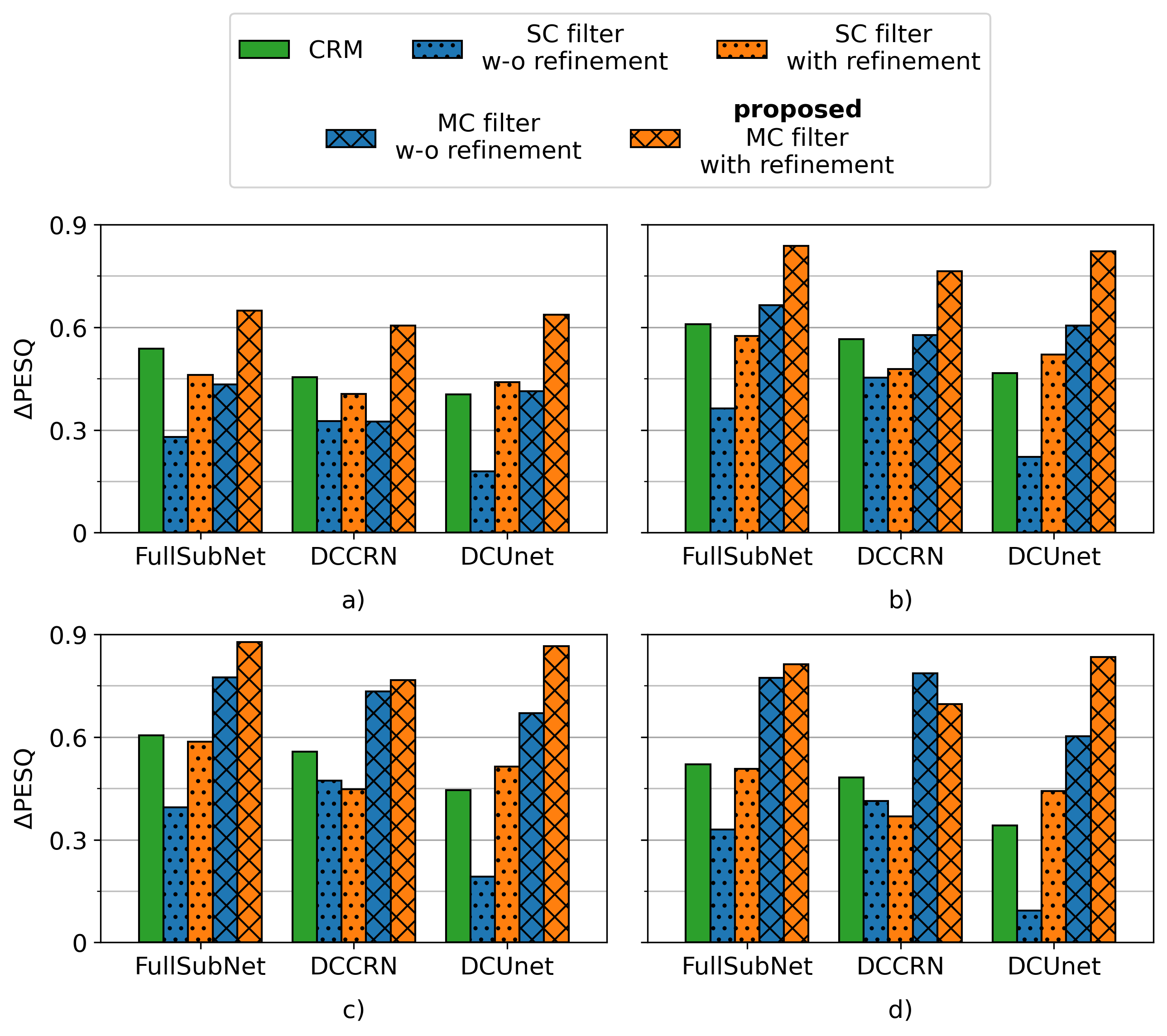}}
\vspace{-8pt}
\caption{Averaged $\Delta PESQ$ for noisy mixtures with input SNR equal to: a) \SI{-5}{\dB}, b) \SI{0}{\dB}, c) \SI{5}{\dB}, and d) \SI{10}{\dB}. The higher value the better result.}
\label{fig:pesq_imp}
\vspace{-2mm}
\end{figure}

Finally, PESQ improvement results depicted in Figure \ref{fig:pesq_imp} for various input SNRs indicate that the largest and consistent improvement in PESQ, and hence the highest enhanced signal quality, can be achieved using the proposed method. Interestingly, these results imply also that (i) multi-channel filtering brings about the highest quality for high SNRs, irrespective of the accuracy of applied masks, and (ii) the refinement procedure enables to improve the real Wiener gain in single-channel processing so that the enhanced output signal is of a similar quality as in the case of single-channel filtering using complex masks.

\section{Conclusions}
\label{sec:conc}

In this paper, we have proposed a noise reduction method in which multi-channel Wiener filtering is performed based on the refined time-frequency masks of any single-channel DNN model. The proposed mask refinement procedure improves the accuracy of preliminary mask estimates using an EM algorithm based on the CGMM model. The results of performed experiments show that the proposed approach increases the accuracy of mask estimates and, as a result, leads to improved noise reduction.

\bibliographystyle{IEEEtran}

\bibliography{mybib}

\end{document}